\documentclass[nohyper,12pt,letterpaper]{JHEP}
\usepackage{epsfig}
\newfont{\frak}{eufm10 scaled 1200}

\newfont{\Bbb}{msbm10 scaled 1200}     %instead of eusb10
\newcommand{\mathbb}[1]{\mbox{\Bbb #1}}
\DeclareSymbolFont{AMSa}{U}{msa}{m}{n}
\DeclareSymbolFont{AMSb}{U}{msb}{m}{n}
\let\Box\relax
\DeclareMathSymbol{\Box}{\mathord}{AMSa}{"03}

\def \eqn#1#2{\begin{equation}#2\label{#1}\end{equation}}

\title{Recurrent Nightmares? : Measurement Theory in de Sitter Space}

\author{T. Banks,\\
NHETC, Rutgers University, and\\
SCIPP, U.C. Santa Cruz\\
E-mail:\email{banks@scipp.ucsc.edu}}

\author{W. Fischler,  S.
Paban, \\
    Department of Physics\\
    University of Texas, Austin, TX 78712\\
E-mail: \email{fischler, 
paban, @physics.utexas.edu}}

\abstract{The idea that asymptotic de Sitter space can be described by a
finite Hilbert Space implies that any quantum measurement has an
irreducible innacuracy.  We argue that this prevents any measurement from
verifying the existence of the Poincare recurrences that occur in the
mathematical formulation of quantum de Sitter (dS) space. It also implies that
the mathematical quantum theory of dS space
is not unique.  There will be many different
Hamiltonians, which give the same results, within the uncertainty in all
possible measurements.}


\preprint{\hepth{} \\RUNHETC-2002-35 \\SCIPP-02/25\\ UTTG-11-02}
\begin{document}

%%%%%%%%%%%%%%%%%%%%%%%%%%%%%%%%%%%%%%%%%%%%%%%%%%%%%%%%%%%%%%%%%%%%%%%%%%%%
%          Table of contents automatic !!!                                 %
%%%%%%%%%%%%%%%%%%%%%%%%%%%%%%%%%%%%%%%%%%%%%%%%%%%%%%%%%%%%%%%%%%%%%%%%%%%%

\section{Introduction}

A remarkable series of recent observations \cite{deBernardis:2001xk}-\cite{Riess:1998cb}( for a recent review see \cite{Primack:2002th}), indicate
that we live in an accelerating universe. It is quite tempting to
speculate that the source of this acceleration is
a small positive cosmological constant. If this is indeed the case, we
would be living in an asymptotic de Sitter (AsdS) space. Some years ago,
we argued \cite{Fischler,Banks:zj} that the quantum theory of such a universe has a
finite dimensional Hilbert space.

Loosely speaking, a quantum version of the principle of general covariance
would seem to suggest that a physical clock must be a quantum observable.
This in turn suggests, that the clocks in an AsdS universe can only
have discrete  "ticks and tocks" and moreover it seems inevitable that
these devices can only measure a finite total time interval.  The spectrum
of any observable in a finite dimensional quantum system is discrete and
bounded.

At the present time, we only have a rigorous understanding of theories
of quantum gravity with asymptotic timelike or null boundaries.  In these
systems, time evolution is governed by asymptotic symmetries, and we can
imagine all clocks to be sitting at infinity.  The measuring devices in
such a universe can be considered to be arbitrarily large and classical,
without having any effect on the scattering\footnote{We use the word
scattering as a shorthand both for the true scattering matrix in
asymptotically flat spacetimes, and the boundary observables of
asymptotically AdS spacetimes.} experiments that they are measuring.
This is not true in an AsdS universe.  The purpose of the present paper
is to address the question of what can actually be measured in such a
universe.

In the absence of a complete proposal for the quantum theory of AsdS
spaces, we rely on the intuition provided by the Wheeler-DeWitt
formulation of quantum gravity to address these questions.  We will find
that the statement that physical clocks are operators in the physical
Hilbert space of a general covariant system, is {\it not} a formal
property of Wheeler-DeWitt quantization. Rather, we are led to a
description of such a system in terms of a collection of different,
more or less conventional, Hamiltonian evolutions on the physical
subspace.  These evolution operators do not in general commute with
each other.  We have previously suggested that this lack of commutativity
is the origin of
Black Hole Complementarity\cite{Susskind:1993if,'tHooft:1995dq} and have generalized this notion to
other spacetimes with horizons\cite{Banks:2001yp}.

We will then argue that in a true quantum theory of cosmological
spacetimes, continuous time evolution is probably an illusion.
Furthermore, by studying actual measurements in AsdS spacetimes,
we will argue that no reliable measurements can be made over arbitrarily
long periods of time.  We suggest that this means that the Poincare
recurrences of the mathematical formalism have no operational meaning 
\cite{Dyson:2002nt,Dyson:2002pf}.
Finally, we argue that, as a consequence of the inevitable imprecision
in measurements, there is no single mathematical theory of quantum AsdS
spacetimes.  Rather, there is a collection of different time evolution
operators, which will make mathematical predictions for all possible
measurements, with differences that are smaller than the fundamental
imprecision of the measurements themselves.  It is only in the limit of
vanishing cosmological constant that precise mathematical statements have
operational meaning. Equivalently, in this limit, there are {\it scaling
observables} (in the sense of the renormalization group) that are
universal, while many of the corrections to scaling are in principle
unobservable.

The plan of this paper is as follows:

In the next section, we briefly review the arguments for a finite
number of physical states in AsdS universes, and introduce a distinction
between localizable states and horizon states. We then review the formal
Wheeler-DeWitt quantization of gravity and its implication for the nature
of clocks. In section III we review
our understanding of quantum measurement theory.  In particular we
emphasize that it depends to a large extent on the locality properties
of quantum field theory.  In AsdS spaces, and on the horizons of black
holes, there are many low energy states that cannot be simultaneously
described by quantum field theory.  While almost classical measuring
devices {\it might} be constructed for other quantum systems with large
numbers of states, we know too little about horizon states to describe
a measurement theory for them, which does not involve a localized
external observer.   Thus, we will restrict our attention to localizable
measurements.   In section IV, we show that very simple estimates
for localizable measurements, imply that the time for a quantum
fluctuation to fuzz out the classical pointer positions of any localized
measuring device in AsdS space is much smaller than the dS recurrence
time.  This is the central point of this paper.  We also identify
a set of measurements in dS space which asymptote to the S-matrix
observables of asymptotically flat space, and explore the relation
between these idealized observables and observations made by a localized
timelike observer.  Finally, we comment on the likelihood that time
evolution in cosmological spacetimes is fundamentally discrete.

We should also mention that other authors have touched on the issue of
the limitations of measurement in asymptotic de Sitter spaces 
\cite{Bousso:2002fi}, \cite{Bousso:2002ju},\cite{Witten:2001kn}.

\section{\bf Infinite aspirations and finite reality}

\subsection{The phase space of AsdS gravity}

Much of the recent literature on dS space is focused on the attempt to
construct gauge invariant meta-observables, which are analogs of the
boundary correlators of the AdS/CFT correspondence.  We believe that
these attempts are misguided, both because these meta-observables
are mathematically ill-defined, and because (and this is a central
message of the present paper) physical measurements in AsdS spaces
have (in principle) limited precision.  In a sense we will try to make
precise in section IV, this imprecision can be viewed as a lack of "gauge
invariance" of real measurements.

One way to approach this problem\footnote{Which is different from the
arguments we have previously given in published work.} is to think of the
phase space of classical gravity; the space of solutions with given
boundary conditions. We are used to thinking about the space of solutions
to a field theory by invoking the Cauchy-Kowalevska (CK) theorem to map it
into the space of initial conditions on a fixed spacelike slice.  We
believe that in General relativity, this theorem is misleading because of
the generic occurrence of singularities.  We have very little knowledge
about global existence theorems for GR.  The CK philosophy leads us to to
view GR as a field theory.  Instead we believe (invoking a conservative
version of the Cosmic Censorship conjecture) that generic solutions with
scattering boundary conditions lead at most to spacelike singularities
cloaked behind black hole horizons.  The Bekenstein-Hawking formula for
black hole entropy, and the parametrization of phase space by scattering
data, then lead us to a {\it holographic} counting of the degrees of
freedom of GR.

Let us apply these ideas to AsdS spaces.  Consider generic boundary
conditions on either of ${\cal I}_{\pm}$.  The present authors (and
probably a number of relativists before them) have conjectured that apart
from a compact submanifold in boundary condition space, such solutions
have a spacelike singularity of Big Bang or Big Crunch type\footnote{G.
Horowitz has informed us that he and N. Itzhaki have found a proof of
this result in certain circumstances.}.  Intuitively\footnote{We work in
global coordinates.  Energy refers to the nonconserved global time translation
operator.  Energy density is covariantly conserved in the dS background
geometry.  We can use it at a good physical parameter to the extent that
the dS background is undisturbed by the perturbation we are studying.},
if we inject too
much energy at the boundary,
the energy density reaches that at which black holes as large as
the spatial sphere form,
before the sphere shrinks below some finite size. We will call this
minimal size the "radius at the singularity".
   The singularities
for different solutions, will occur at different values of the radius at the
singularity,
because black holes of different sizes will be formed.  Thus, we believe
that, apart from the compact subspace of the naive asymptotic phase
space, for which global solutions are nonsingular or contain black holes
smaller than the minimal dS radius,
the existence of solutions depends on the resolution of
singularities.  Furthermore it seems clear that solutions in which the
singularity occurs at different radii should not be considered part of
the same theory\footnote{We are not claiming that all of these solutions
describe sensible quantum theories, only that , if they do, then solutions
with different radii at the singularity, correspond to different quantum
systems.}  The consequence of this point of view is that the space of
asymptotically dS solutions of GR ((generically)
either in the past or future, but not
both) breaks up into compact subspaces.  Each such subspace, upon
quantization , should lead to a finite number of physical quantum states.

One of these quantum theories corresponds to solutions, which are AsdS
both in the past and the future.  We think of this as the quantum theory
of idealized dS space.  Among the future AsdS spaces with a Big Bang
singularity, we imagine that there is at least one subspace, which defines
a sensible quantum theory that describes the Universe we observe.
In the rest of this paper we will explore the idealized dS quantum
theory, since the issues we address have nothing to do with the Big Bang.

\subsection{Wheeler DeWitt quantization}

The residual gauge invariances of the synchronous gauge in General Relativity
are the freedom of choice of the initial spatial slice, and spatial
reparametrizations of that slice.  Correspondingly, there are two classes of
constraints:

\eqn{wda}{{\cal H}_i ({\bf x}) = 0}
\eqn{wdb}{{\cal H} ({\bf x}) = 0.}

The first set merely state that physical quantities are diffeomorphism
invariant on the spatial slice.  They are kinematical, and easy to solve.
The second constraint is the Wheeler-DeWitt equation, which is dynamical.

The essential point we want to make is simplified, if we further choose the
gauge, so that only a global reparametrization of time is allowed.   Then
the Wheeler-DeWitt constraint becomes, upon formal canonical quantization,
  a single, hyperbolic differential equation:

\eqn{simpwd}{[G^{ij} (\phi)  \partial_i \partial_j + U(\phi)]\Psi = 0  }

The $\phi$ are the canonical coordinates of the system and the DeWitt metric
$G_{ij}$ has signature $(1,N)$. The wave function $\Psi$ might be a finite
dimensional vector of wave functions, and the potential $U$ might be a matrix
in this finite dimensional space.
  The simplest example of a time
reparametrization invariant system is the relativistic particle.  In this
case the Wheeler-DeWitt equation is just the Klein-Gordon equation.
String theory teaches us that in order for this equation to make sense at the
quantum level, one must generally include contributions from Fadeev-Popov
ghosts in the Wheeler-DeWitt operator.   However, it is not our intention
to propose Wheeler-DeWitt quantization as a rigorous approach to a realistic
theory of quantum gravity, so we will ignore this technicality.

The space of wave functions on which the Wheeler-DeWitt operator acts is not
a Hilbert space.  It is only the solution space of the equation which
is to be interpreted as the physical Hilbert space of the system.
Solutions of the WD equation are parametrized, according to the CK theorem
by the values of $\Psi$ and its normal derivative, {\it on a spacelike surface
in the space of variables $\phi^i$}.   In a correct quantum theory,
this solution space, or some subspace of it,
  will carry a positive definite inner product. Note that
the solution of the WD equation will define a one parameter set of
vectors in this physical Hilbert space.  The inner product should be
defined in such a way that it does not depend on which spacelike slice
in field space it is evaluated on.  Then the one parameter set of vectors
will be related by unitary evolution in the Hilbert space.

One often tries to give an intuitive description of this mathematical procedure
by arguing that the physical meaning of time reparametrization is that
clocks must be chosen to be physical objects.  The choice of gauge, which
reduces the full-blown WD equation to a hyperbolic PDE with only one negative
eigenvalue of $G_{ij}$ is ``a choice of physical clock''.  Note however that
the clock variable is not a quantum variable in the physical Hilbert space.
Rather, it defines a one parameter family of states in the physical Hilbert
space, much like a classical time variable.

In fact, the problem of constructing a positive metric subspace of WD solution
space can only be solved in a general manner in the semiclassical
approximation\cite{Banks:1984np}.  Here one chooses a class of solutions
of the WD equation of the form (the justification for this form depends on
the details of the WD equation)
\eqn{wdwkb}{\Psi \sim e^{i {S(t) \over \epsilon}} \psi(t, \phi^k ),}
where $\epsilon$ is a parameter whose small size justifies the WKB
approximation. $t$ is a particular combination of the coordinates $\phi^i$,
which parametrizes a spacelike slicing of a particular classical solution
of the system
$\psi$ satisfies a time dependent Schrodinger equation
\eqn{schr}{i \partial_t \psi = H(t) \psi}
and the solutions of this equation have a well known time independent
, positive, scalar product, if $H(t)$ is Hermitian.

In general, for a given classical solution of the system, there may be
many ways to choose the time slicing.  Each of these defines a different
unitary evolution on the same Hilbert space.  There is no reason for these
different evolution operators to commute with each other, at any time.
This leads to a quantum complementarity between the observations of
observers using different classical coordinate systems.  In a quantized
theory with general coordinate invariance there is no reason to expect
to be able to discuss simultaneously, the observations of two observers
using different coordinate systems.  The authors have argued that this
is the origin of Black Hole Complementarity\cite{Banks:2001yp}.

Beyond the semiclassical approximation, there are apparent problems with
the straightfoward WD approach.  The Klein-Gordon example provides some
insight into these problems.  In that context they have to do with the
mixing of positive and negative frequency modes (in general the scalar
product is positive only on one of these subspaces) and the necessity
for second quantization.  There is no satisfactory resolution of these
problems.  For this and other reasons
(principally the spatially local parametrization of phase space - which
seems to us incompatible with holography), we do not believe that the WD
equation provides a correct definition of a quantized theory of gravity.

However, it seems extremely plausible that certain features of WD quantization
will survive in a well defined quantum theory of gravity.  The existence
of many equivalent but complementary time evolution operators (which generally
do not have a time independent Hamiltonian) would seem to be a general
consequence of combining quantum mechanics with general covariance.
We will argue later that there may be contexts in which time evolution
is discrete rather than continuous, but that does not change the general
structure in a drastic way.

\section{\it Measurement theory and locality}

At the heart of the quantum theory of measurement is Von Neumann's
theorem that unitary evolution can take a product state $$|A_0 > \otimes
\sum c_s |s>$$ into an entangled state $$\sum c_s |A_s > \otimes |s>$$.
We think of $|s>$ as a complete orthonormal basis of the quantum
system to be measured, while the states $|A_i >$ are states of the
apparatus.  Von Neumann shows that the apparatus states in the entangled
state can be chosen orthonormal. Indeed, unitary evolution can take any
state in the tensor product Hilbert space into any other with the same norm.
If the $|A_s >$ are orthonormal, expectation values
of an arbitrary system operator $O_s$ in the entangled state, are given
by
\eqn{classexp}{<O_s> = \sum |c_s|^2 <s| O_s |s>}
This is the expression for the classical expectation value of a quantity
in a probability distribution $P_s = |c_s|^2$.  Quantum interference
has disappeared.  We say that we have measured those operators of the system,
which are diagonal in the $|s>$ basis.

The problem with this argument is that the apparatus itself is a quantum
system and we can imagine reexpressing this same entangled state in terms
of another orthonormal basis for apparatus states.  Then the same
measurement can be interpreted as measuring something else.  In the case
where the $c_s$ are all equal we can interpret the same time evolution as
a measurement of operators which do not commute with the $|s>$ projectors.
Very clear discussions of this problem can be found, for example, in
\cite{Everett:hd,Zurek:xq}.

Much recent literature on this problem has focused on the resolution of
this ambiguity, "the choice of pointer basis", by environmentally induced
superselection, {\it i.e.} the interaction of the apparatus with an
unobserved environment.  It is often stated that the size of the
apparatus cannot by itself resolve this ambiguity.

Our own understanding of the measurement problem is somewhat different
, and does depend on the size of the Hilbert space of the apparatus.  At
present it also depends on a good approximate description of the physics
of the apparatus by (perhaps cut-off) local field theory.  As orientation
for this discussion let us remind the reader of the discussion of
spontaneous symmetry breaking in field theory (or more general occurrences
of moduli spaces of vacua).  In particular we want to draw attention
to the well known fact that one is {\it not} allowed to make
superpositions of states which have different constant expectation values
of a scalar field.  The essential reason for this is locality.  The
Hilbert space of a field theory is constructed by acting with local
operators on a single Poincare invariant vacuum state.  States with
different Poincare invariant expectation values cannot be related to each
other by local operations.

This principle is valid for cutoff field theories, and it even has
implications for large but finite systems.  In the latter, states with
different expectation values can communicate with each other via local
operations, but their overlaps are of order $e^{- V}$ where $V$ is the
volume of the system in cutoff units.   Furthermore, although in systems
with a symmetry, states with expectation values of symmetry breaking
fields are not energy eigenstates, the quantum tunneling amplitudes
that convert such an initial state into a properly symmetric eigenstate
are exponentially small.  Thus, in a large but finite system there are
long lived states that have expectation values of quantities that are
macroscopically distinguishable.  In fact such states are members of huge
subspaces of the Hilbert space which have almost the same value for
these macroscopic operators.  The expectation values are thus robust
in the sense that many local operations on the system do not change them
(or change them by inverse powers of $V$).  Of course, local operations
that dump huge amounts of energy onto a finite system {\it can} change
macroscopic expectation values\footnote{For example, inserting a tiny, but
macroscopic, amount of antimatter into any piece of laboratory equipment,
will disrupt its ability to measure anything.}, but there is a wide range of
perturbations which do not.

Our understanding of measurement theory is based on such macroscopic
observables in local field theory.  For us, a measuring device must be
huge in microscopic terms, and have many more states than the system it
is measuring.  It must have macro-observables like the expectation values
of scalars in field theory, which take on almost the same value in a
finite fraction of the states of the apparatus.  The overlap between
states with different values of the macro-observables must be very small,
and tunable to zero as the size of the apparatus goes to infinity.
Large classes of perturbations of the apparatus must leave the subspaces
with fixed macroobservables invariant, with very high probability.
Cutoff local field theory for a finite volume system provides
us with many explicit examples of systems with these properties.  There
may be other examples, but to our knowledge there does not exist a
systematic theory of nonlocal measuring devices.

Notice several points about our description:  the exact orthogonality
of the $A_s>$ assumed in Von Neumann's discussion is unnecessary.  It is
sufficient that the overlaps of different pointer states are exponentially
small at larger pointer size.  We do not then recover an idealized measurement
as expressed in the axioms of the Copenhagen interpretation.  The
expectation values in the entangled state do not reproduce the classical
probability sum rules exactly but only up to exponentially small
corrections.  The volume that appears in these
exponentials is that of the pointer rather than of the whole apparatus.
Finally, these remarks imply that for a finite apparatus the whole
concept of measurement is only an approximation.  Not only are the
classical probabilities achieved only in an approximate manner, but the
measurements are not exactly robust.  Given a long enough time, quantum
fluctuations of the pointer will nullify the utility of the apparatus as
a measuring device.  Ideal Copenhagen measurements are only achieved in the
limit of infinite apparatus.

The above discussion ignores the interaction of the apparatus with its
environment - that is the effects which are taken into account for
environmentally induced superselection.  These are important for ordinary
experiments, but not, we believe, for the idealized "maximally accurate''
experiments in dS space that we will study in the next section.  That is,
while we accept the fact that environmentally induced superselection may
be quantitatively important in understanding the behavior of many everyday
measuring devices, we do not believe that environmental decoherence is
necessary to the construction of a sensible measurement theory.  In particular,
in the next section we will discuss idealized measurements in dS space
in situations where the measuring apparatus does not seem to have much of
an environment.   We claim that large, approximately local, devices are capable
of making almost ideal quantum measurements without the benefit of an
environment.  A measurement consists of entangling a complete set of basis
states of the measured system with macroscopic, approximately orthogonal,
states of the apparatus.  In the limit of infinite apparatus size,
macroscopic states become orthogonal superselection sectors.  It is possible
that other large systems can also act as measuring devices.  In particular,
black hole and cosmological horizons are repositories of large numbers of
quantum states.  It is possible, but beyond our current capabilities to
prove, that horizons or pieces of them, can act as measuring devices.

\section{\it Measuring de Sitter}

In any accurate measurement, one wants to minimize the effect of the apparatus
on the measured system.  We want to correlate the states of the system
prepared in some experiment, with macroscopic apparatus states, without having
the apparatus interfere with the experiment.   In a theory of gravity this
is difficult because any apparatus has a long range, unscreenable, interaction
with any experiment.  The Newtonian estimate of the strength of this
interaction is ${G_N M \over R}$, where $M$ is the mass of the apparatus,
and $R$ is its minimal distance from the experiment.  We could try to minimize
this interaction by making $M$ very small, but this interferes with our ability
to make the large, almost classical, devices required by measurement theory.
The only alternative, is to make the distance $R$ very large.

This simple estimate is a powerful hint that the only well defined mathematical
observables in a quantum theory of gravity should be scattering data.
That this is indeed the case in string theory is well known, and has led to
a lot of angst about the appropriate observables in dS space.  Our approach to
this problem\cite{Banks:2001yp}
will be to rely on the idea that asymptotically flat space is
a limit of dS space.  Thus, there should be mathematical quantities in any
quantum theory of dS space, which approach the flat space S-matrix as the
cosmological constant goes to zero.  As in any limiting situation, there may
be ambiguities in what these quantities are for finite values of the
cosmological constant.  Our strategy in this paper
will be to find a physical definition of
these approximate S-matrix quantities.  We will find that they do not
correspond to observations that can be made by local observers in dS space.
We will then try to elucidate the connection between these two kinds of
measurement.  Throughout our discussion, we will ignore the special
complications of four dimensions, where the flat space S-matrix does not
exist because of the infrared problems of graviton bremstrahlung.

Measurement theory in dS space is caught in a web of contradictory
requirements.  To minimize quantum fluctuations in our apparatus,
we must make it large.
This requires us to move it far away from the experiment, but the
existence of a cosmological horizon makes it difficult to do this.
A related problem is the putative existence of only a finite number of states
in dS space.  Any measurement in such a finite system consists of a split of
the states into those of a ``system'' and those of an apparatus.  A bound
on the total number of states means that the apparatus cannot approach the
classical behavior required of an ideal measuring apparatus.

There is a further problem, which we encounter if we attempt to make our
apparatus stationary with respect to the frame of the experiment.
It then becomes impossible to move it far away.  If we consider a sequence
of stationary measuring devices, closer and closer to the horizon of the
experiment, the later devices in the sequence find themselves in a hotter
and hotter environment.  Eventually they are destroyed by Hawking radiation.

This indicates that the best experiments one can do in dS space are those
performed by machines freely falling with respect to the experiment, machines
that will eventually go outside the experiment's cosmological horizon.  We
consider experiments whose ``active radius'' remains finite as the dS radius
goes to infinity.   By active radius, we mean the radius outside of which
the experiment can (with some accuracy) be said to consist of freely moving
particles in Minkowski spacetime.   Of course, if we really want to get precise
answers, we have to let the active radius go to infinity with the dS radius.

We consider machine trajectories whose closest approach to the experiment goes
to infinity as a power, $R_{dS}^{\alpha};\ \ \alpha < 1$,
of the dS radius.  Such machines
can be taken to have a mass $\mu$, which scales like $\mu \sim R_{dS}^{\beta}$.
As long as $\beta - \alpha < 0$, the gravitational effect of the detector
on the experiment, will vanish as the dS radius goes to infinity.
Thus, we can make very massive, freely falling machines, whose gravitational
effect on the experiment is negligible.

Further constraints come from the requirement that our machines are
approximately described by local field theory.  We emphasize that this
requirement would not be fundamental if we could learn how to manipulate
the internal states of black holes or the cosmological horizon to make
measuring devices.  If we rely on ``current technology'' then the best
we can do is to use the high energy density of states of a field theory
to make our measuring devices.
These then have an energy and entropy density,
\eqn{enden}{\rho \sim T^4 ; \ \ \sigma \sim T^3}
The requirement that a machine made out of such conformal stuff not collapse
into a black hole, restricts its total entropy to scale like $R^{3/2}$ where
$R$ is the linear size of the (roughly isotropic) machine. This is 
the same power law
that one obtains for the entropy of a starlike solution of the 
Tolman-Oppenheimer-Volkov
(TOV) equations \cite{bondi, wald, Banks:2002fj} We have seen above that
$R$ can scale at most like a power of
$R_{dS}$ less than
$\alpha$.

Now the crucial point is that the tunneling amplitudes between macroscopic
state of the apparatus are of order $N_{ap}^{-p}$ where $p$ is an exponent
of order one and $N_{ap}$ the number of states of the apparatus.   This is a
consequence of the fact that if we write the Hilbert space for a local
system as a tensor product of spaces of localized states, then by definition
two macrostates different from each other by an amount of order one in each
of a finite fraction of the factors.  The total number of states is
\eqn{totnum}{N_{ap} = n^k}
where $n$ is the number of local states and $k$ the number of localized
factors.  The overlap between two macrostates is
\eqn{overlap}{<\Psi_1 | \Psi_2 > \sim \prod_{i=1\ldots ak} o_i,}
where $a$ is a fraction of order one and each $o_i$ is a local overlap, a
number less than one, but of order one.  This is of order 
$N_{ap}^{-p}$, as claimed.

Putting together these results, we find that the tunneling time between
different macrostates representing different pointer positions on our
measuring device, is of order $e^{c R^{3/2}}$, where $R$ is the size of
the device.   It is clear that $R$ can scale at most like some power
, from the scaling law of the TOV equations, the size scales like the 
mass: $R \sim
R_{dS}^{\beta}$, where as shown above,
$\beta <
\alpha <1$ if we want to avoid interference between the device and 
the experiment. 
{\it Thus, the tunneling time between macrostates of any sensible 
local measuring
apparatus in dS space, is much shorter than the Poincare recurrence 
time.}  Indeed, for
a dS radius corresponding to the observed acceleration of the 
universe, the recurrence
time is so long that it is essentially the same number expressed in seconds
as it is in units of the tunneling time of the largest possible apparatus
in dS space.   As a consequence, no local measurement (which, in our current
state of ignorance of horizon physics means no measurement we know how to
describe) can be made over time periods remotely comparable to the recurrence
time.  A mathematical description of dS space may contain the phenomenon of
Poincare recurrence, but there is no apparent way to test this mathematical
conclusion by experiment.  Quantum fluctuations in the measuring device
will wipe out all memory of measurements long before the recurrence occurs.

Environmentally induced decoherence could lengthen the time period of utility
of a quantum measuring device.  This is certainly an important quantitative
effect for normal laboratory equipment.  Indeed, it has been claimed that
the decoherence due to interaction with the solar wind at the orbit of
Saturn is necessary to explain the classical rotational behavior of one of its
satellites, Hyperion \cite{Zurek:1994bn}. However, the experiments we are
contemplating take place in
an environment much more rarefied than interplanetary space.   We are
discussing observables in dS space that approach the S-matrix of an
asymptotically flat spacetime in the limit of vanishing cosmological constant.
The latter describes the behavior of finite numbers of particles in an
infinite empty universe.  There is no environment around to help the idealized
``S-matrix meters'' decohere, and we must rely on the superselection inherent
in the behavior of these large local machines to perform measurements.

Finally we note that even if we could build measuring devices from the
cosmological horizon states, we would find that the tunneling time between
pointer macrostates of the horizon would be much smaller than the recurrence
time, since such states could at most carry a finite fraction of the total
entropy.

It is obvious that, given the size of the cosmological constant implied by
observations, these restrictions on measurement have no practical content.
The technological restrictions implied by our own local nature will, forever,
make the measurements we can actually do much less accurate than these
bounds would allow them to be.  The significance of these bounds is to
call into question the utility of discussing the recurrence time scale and
to show that no mathematical quantum theory of AsdS space can be unique.

Indeed, it is clear that there must be many Hamiltonian descriptions of the
same system, whose predictions are sufficiently similar that no conceivable
experiment could distinguish them.  The complete set of observables in
a quantum system with a finite number of states, can only be measured by
an infinite external measuring device.   Any self measurement in such a system
requires a split of the states into observer and observed, and an intrinsic
uncertainty in measurements as a consequence of the quantum fluctuations of
the measuring device.  If we cannot measure all the mathematical consequences
of a given definition of observables, then there will be many different
mathematical theories which have the same consequences for all measurements.
We view this as a kind of gauge equivalence, related to the infamous Problem
of Time in the WD quantization of gravity.  The Problem of Time in a closed
universe is the absence of a preferred definition of time evolution.  Different
choices of time evolution are formally related by gauge invariance.
But there are no gauge invariant operators.   Our discussion of measurement
theory in a finite system, leads to a similar conclusion: different choices
of Hamiltonian with indistinguishable consequences for measurements\footnote{
The Problem of Time is more general and involves gauge equivalence between
observers whose experience {\it is} measurably different.  For example, the
principle of Cosmological Complementarity\cite{Banks:2001yp}, which we related
to the Problem of Time, says that in the late stages of our AsdS universe,
the observations of people in the Sombrero galaxy are gauge equivalent to
our own.  Each of us views his own physics as local, and his erstwhile
companion's physics as a quantum fluctuation on the event horizon. }.

\subsection{Local observations}

The measurements we have been discussing so far are not realistic measurements
from the point of view of a local observer in dS space.  The machines that
measure initial and final states of particles at different angles
in our scattering experiment are outside each other's event horizons.  There
is, until we take the $\Lambda \rightarrow 0$ limit, no single observer who
can collect all of this information.  We now want to discuss actual
measurements by an observer gravitationally (or otherwise) bound to the
experimental apparatus (the accelerator, not the detectors).   As far as
we have been able to tell, the most accurate and nondestructive measurements
this observer can make on his experiment consists of receiving information
from the free falling ``S-matrix meters'' that we have discussed previously.
That is, one must equip those devices with a signaling mechanism, which
can send photons or other messenger particles back to the observer at the
center of the experiment's horizon volume.

It is clear that the local observer can garner {\it less} information than
is available to the freely falling machines.   There are two separate effects
here.  First of all, as the machines approach the horizon, and their
disturbance of the experiment goes to zero, the photons they emit are
redshifted.  For any fixed lower frequency cutoff on the local observer's
photon detection system, there will come a time when signals from the
freely falling machine are undetectable.   Furthermore, the photons sent
to the center from the freely falling machine are {\it accelerated}.  Thus,
they will experience interaction with the Hawking radiation.  As the machine
approaches the horizon, this radiation gets hotter and denser.   These random
interactions will tend to destroy the information contained in the signal
sent to the local observer.   Thus, we can understand, in terms of relation
between the S-matrix meters and local observations, the fact that the local
observer finds the universe to be in a mixed state, while the S-matrix
observations retain their quantum mechanical purity.

A final constraint on the accuracy of measurements made by a local observer
comes from the fact that he is immersed in thermal bath at the dS temperature.
Although this temperature scales like $1/R_{dS}$, the probability for a
catastrophic burst of photons or the nucleation of a black hole which would
engulf him is nonzero. The timescale between such events is,
again, much less than the recurrence time.  This phenomenon also occurs
for the freely falling S-matrix meters.  However, since their size is
taken to scale to infinity with the dS radius, the effect can be made small
in the limit of infinite radius.  The predictions for the limiting Minkowski
theory will be precise quantum measurements.

\subsection{Discrete time?}

This subsection is not strongly correlated with the rest of the article.
We include it because our original speculations about physical clocks suggested
a discrete and finite time evolution.  We have already seen that, although
the mathematical formalism appears to be compatible with arbitrarily long
times, we cannot set up measuring devices within the system that could
measure correlations over (much less than) a Poincare recurrence time.

It is certainly true that the static Hamiltonian in dS space has a maximal
eigenvalue, corresponding to the Nariai black hole.  Standard discussions
of the energy time uncertainty relation then imply that we cannot set up
time measurements of intervals shorter than the inverse of this energy.
In a certain sense then, our intuition is correct.

A more formal argument that time evolution is discrete comes from our proposal
for describing quantum mechanical Big Bang cosmologies\cite{Banks:2001yp}
\cite{tbmill}.  Here, the results of all possible experiments done by a
local observer are described by a sequence of Hilbert spaces of
(multiplicatively) increasing, finite, dimension.
A given Hilbert space describes all
experiments that could have been done in the causal past of a given point
in spacetime.  Time evolution along a particular timelike trajectory
must be compatible with the state of the system breaking up into tensor
factors at earlier times, with one factor describing all data that could
have been accumulated inside of the particle horizon at a given point on the
trajectory. Since Hilbert space dimensions are discrete, there is a natural
discrete breakup of the time interval.

If we look at typical FRW cosmologies, the Fischler Susskind Bousso (FSB) area
of the particle horizon scales like a power of the cosmological time.  Thus,
the minimal increase in area, corresponds to a smaller and smaller increase
in cosmological time, as time goes by.  The time resolution becomes finer
and finer, and goes to zero if the particle horizon eventually becomes
infinite\footnote{Note that this is in perfect accord with arguments based
on the energy-time uncertainty relation.  An observer with a larger particle
horizon can access states of higher energy and therefore has a finer time
resolution.}.  However, in AsdS space, there is an ultimate limit to this
increase in resolution, because the particle horizon ceases its growth.

\section{Conclusions}

We have argued that a universe which
asymptotically approaches dS space presents a
serious challenge as to what is observable within that universe and
how precise the measurement of such an observable can ever be.

We have shown that the essential limitation arises from the existence 
of an event
horizon and the ability to describe AsdS space by a finite 
dimensional Hilbert space.
In a nutshell, the necessity to separate the measuring device from 
the experiment
such as to minimize the gravitational interaction, as well as 
availability of only a
finite number of states to build detection devices, limits in principle the
time period during which a measurement is "reasonably" sharp. The time after
which the pointers on any conceivable local measuring device suffer
uncontrollable quantum fluctuations is much shorter than the Poincare
recurrence time.

One of the consequences of these considerations not discussed in the 
text is that an
inflationary epoch, for example due to a slowly rolling inflaton, in an
AsdS universe is
limited to be much shorter than a recurrence time.
The reasons for such limitation is
again the availability of a finite number of states to describe the dynamics
of a scalar field.  The quantum fluctuations of the inflaton field will
wipe out the classical slow roll on a time scale shorter than the recurrence
time.  Therefore, although one can write down classical cosmologies in which
an arbitrary number of e-foldings of inflation are followed by an AsdS era with
any value of the cosmological constant smaller than the inflationary vacuum
energy, many of these models will be ruined by quantum fluctuations.

It may be possible to model such dynamics by imposing a q-deformed algebra on
  the creation and annihilation operators of the inflaton, where the deformation
parameter is the N-th root of unity and N is an integer commensurate 
with the finite
number of the states.  It seems clear that, for the value of the 
cosmological constant
consistent with observations, the constraints on  inflationary models 
are very mild.
Even $10^{10}$ e-foldings of conventional slow roll inflation would 
be compatible with
the constraints of {\it our} AsdS universe.

Our considerations suggest that any mathematical description of AsdS space
is of limited utility.  That is, there will be many different
mathematical theories compatible with all possible experimental measurements,
within the fundamental limitations on experimental accuracy.  Furthermore,
although a mathematical theory of AsdS space may describe arbitrarily long
time scales, the recurrence time puts an upper bound on the operational
meaning of any such description.  Within the limits of ``current technology'',
in which measurement theory is based on local devices, the actual operational
limits on measurement time scales are dramatically smaller than the
recurrence time.   Only if one could learn to exploit and manipulate
properties of horizon microstates could one approach measurements that would
be robust over a recurrence time.

Separate arguments suggest a fundamental limit on the resolution of time
measurements in an AsdS space.  

It is possible to take a pessimistic view of these limitations.  In some
sense they mean that if spacetime is AsdS, then the ideal goal of the
theoretical physicist {\it viz.}, to obtain a mathematically precise Theory
of Everything is unattainable.   One might be tempted to view this as
a philosophical argument against the interpretation of current observations
in terms of a cosmological constant.  We reject such pessimism and its
associated philosophical prejudices.  In fact, the idealistic goal of
a mathematical Theory of Everything seems a bit naive to us.  Physics was
developed as a way of explaining the behavior of isolated systems.  The idea
of experiments and ideally precise measurements presumes a precise, or
at least infinitely refinable, separation between observer and observed.
Quantum mechanics, which must (in our view) be interpreted in terms of
infinitely large measuring devices, makes this idealization even more
difficult to achieve.  It seems almost inevitable that these mathematical
idealizations should fail when applied to the whole universe, at least if
it has a finite number of states. 

The question of whether the universe has a finite or infinite number of
states should be answered by experiment rather than philosophical prejudice
\footnote{The annoying question of what fixes the number of states if it is finite,
cannot be answered in an experimentally testable way.  Perhaps a sufficiently
elegant theoretical answer to this question would calm our uneasiness
about this point.}
.
Classically, one can build models of scalar fields with metastable
minima, which can mimic an arbitrary cosmological constant for long
periods of time, but then relax to an infinite FRW future. 
In such models, it makes sense to talk about times much longer than
the Poincare recurrence time of the temporarily dS universe.
Indeed, this is quite analogous to our interpretation of observations
of the Cosmic Microwave Background in terms of super-horizon size fluctuations
of an inflationary universe. If what we are experiencing today is
a second period of temporary inflation, it is
not clear that measurements made by local observers prior to the end of
inflation will be able to distinguish such a model from a true cosmological
constant.  Such local measurements would be subject to the constraints
discussed in this paper.  It would appear that only observers living
after the Second Inflationary Era ends could verify the existence of
Poincare recurrences.

However, if the conjecture of \cite{Banks:zj}, relating supersymmetry breaking
to the cosmological
constant, and the existence of a true cosmological horizon with a fixed
finite number of states is correct, then we will be able to 
interpret the
measurement of superpartner masses as a verification of the fact that the
universe is finite.  Such a verification would imply that a mathematical
Theory of Everything did not exist and would signal that we have reached
the limits of theoretical physics.  Given the apparent value of the
cosmological constant, these limitations are of no practical relevance,
nor will they ever be.  Philosophical reservations about such a view of
the world seem to us to have little relevance to the real enterprise of
physics.  This has always been to give a simple explanation of the
regularities {\it we} see in nature.  Even with an extremely optimistic
estimate of what our observational capabilities might be in the distant
future, there is no danger that the fundamental limits to observation in
AsdS spacetimes will ever effect the outcome of a real measurement.

%=========================================================================
\acknowledgments
We would like to thank Edoardo di Napoli, Amir Kashani-Poor, Hong 
Liu, Bob McNees,
Jay Park and Philippe Pouliot for useful discussions.

The research of W. F. and S. P. is based upon work supported by the 
National Science
Foundation under Grant No. 0071512. The research of T.B. was supported
in part by DOE grant number DE-FG03-92ER40689.

%=========================================================================

\end{document}